\documentclass[12pt]{iopart}

\def\be{\begin{equation}}
\def\ee{\end{equation}}

\def\te{\end{equation}}
\def\bea{\begin{eqnarray}}
\def\nn{\nonumber\\}
\def\tea{\end{eqnarray}}
\def\eea{\end{eqnarray}}

\begin{document}

\topical[Chaos, decoherence and quantum cosmology]{Chaos, decoherence and quantum cosmology}

\author{Esteban Calzetta}

\address{$^2$ Physics Department, Buenos Aires University, Ciudad Universitaria, Pabell\'on I, Buenos Aires, 1428, Argentina and IFIBA-CONICET}

\ead{calzetta@df.uba.ar}\label{} 
\begin{abstract}
In this topical review we discuss the connections between chaos, decoherence and quantum cosmology. We understand chaos as classical chaos in systems with a finite number of degrees of freedom, decoherence as environment induced decoherence and quantum cosmology as the theory of the Wheeler - DeWitt equation or else the consistent history formulation thereof, first in mini super spaces and later through its extension to midi super spaces. The overall conclusion is that consideration of decoherence is necessary (and probably sufficient) to sustain an interpretation of quantum cosmology based on the Wave function of the Universe adopting a Wentzel - Kramers - Brillouin form for large Universes, but a definitive account of the semiclassical transition in classically chaotic cosmological models is not available in the literature yet.
\end{abstract}

\maketitle

\section{Introduction}

 The goal of this topical review is to explore the connections between the concepts of chaos, decoherence and quantum cosmology. Each of these has a long history and development independently of the other two, and the relationships between chaos and decoherence are varied and deep. When it comes to their relationship to quantum cosmology, on the other hand, it is fair to say that the available literature barely scratches the surface of the subject. Therefore our goal is not only to review what has been done to date, but also to show areas which remain underdeveloped, sometimes in spite of the fact that the necessary tools are already available in non-quantum cosmological contexts.

Two warnings to the reader are called for. First, we shall make no attempt to be systematic in our presentation of chaos \cite{ArnAve68,BalVor86,Berr83,Chir79,ReiZhe87,Shie74} and decoherence \cite{GJKKSZ03,Schl07,Weis93}, we shall only pinpoint those aspects of these subjects which have been found to be relevant to quantum cosmology. Most importantly, we shall neither attempt a complete presentation of quantum cosmology itself \cite{Bojo11a,Bojo11b,Kief07,Varg10a,Varg10b} . For the purposes of this review, quantum cosmology means the theory of the Wheeler-DeWitt equation restricted to Friedmann - Robertson - Walker mini super spaces (to be defined below), later enlarged to midi super spaces by including also linearized primordial fluctuations (whereby the former mini super spaces come to be regarded as quantum open systems). We have made this choice not because we think this is the most important or promising brand of quantum cosmology (if asked point blank, we would rather give that award to the loop quantization program \cite{Asht91,AshSin11}) but because that is the framework in which the connections of quantum cosmology to chaos and decoherence have been most systematically addressed, and where the language remains closest to the other area where cosmology and quantum physics intertwine, namely, the generation of primordial fluctuations during inflation \cite{CalHu08,Mukh05,MuFeBr92} .

Let us try to put the above choice of subject matter in perspective.

It is certainly not coincidence that after a strong paradigm of quantum field theory based on the renormalization program emerged in the fifties and early sixties, people started to worry about extending it to other areas of physics, such as condensed matter theory \cite{DomMar64a,DomMar64b,Fish74,Schw60}, turbulence \cite{Krai59,Krai61}  and gravity \cite{Feyn63,Feyn95}. In the case of gravity, which is a constrained system development had to await for the unraveling of the Hamiltonian structure of the theory \cite{Dira58b}. Once this was accomplished by Arnowitt, Deser and Misner (ADM)\cite{ArDeMi62,MiThWh73}, the road was open for a systematic quantization of gravity on the lines of Dirac's quantization \cite{Dira50,Dira58a,Dira64,HaReTe76}. This program was undertaken by Wheeler, DeWitt and collaborators, and resulted in the formulation of the Wheeler - DeWitt equation, which is the cornerstone of our subject \cite{DeW67a,DeW67b,Whee68,Kuch73,Kuch81}.

This first wave of quantum cosmology research stumbled upon the problem of actually finding solutions to the constraints of the theory (namely, states). Rather than making no progress at all, Misner and others hit upon the idea of working on (very) restricted models where the momentum constraints where trivial, and only the Hamiltonian constraint, or Wheeler - DeWitt equation, remained as dynamical law \cite{Berg74,Berg93a,Misn69,Misn70,Misn72,Misn73,Ryan72,RyaShe75}. At first, these so-called mini super space models were offered as just mathematically consistent models which shared one important property (time-reparametrization invariance) with full quantum cosmology, and could therefore be of use to explore outstanding issues such as the emergence of time and the correct interpretation of the theory. Eventually mini super spaces came to be regarded as all that and also as some kind of Born - Oppenheimer approximation to the full theory, so that the study of minisuperspaces was directly relevant to cosmology \cite{KucRya86,KucRya89}. The problem of actually solving the constraints of the theory was reopened much later by the loop quantization program, which we shall not discuss here \cite{Asht91,AshSin11}.

At about this point the program developed by Wheeler, Misner and their associates merged with a powerful current coming from the (then) Soviet Union, where Landau, Belinskii, Lifshitz, Khalatnikov and their collaborators had made foundational work on the fluctuations of Friedmann - Robertson - Walker cosmologies, and described the homogeneous Bianchi-type cosmologies, including  the notorious Bianchi type IX model  \cite{Beli92,Beli09,BeKhLi70,Kame10,KLKSS85,KhaKam98,KhaKam08,LanLif75}. In time, Bianchi IX came to play a role, with respect to cosmology and chaos, not unlike the one played by the Schwarzschild Black Hole with respect to gravity and thermodynamics.

A third moment in the development of quantum cosmology came when the emphasis shifted from the analysis of the Wheeler - DeWitt equation in general, to the identification of which particular solution to that equation described the state of our Universe, and the correct way to derive cosmological predictions from it. Here the foundational work is no doubt that of Hartle and Hawking, which brought the formalism of euclidean path integrals to bear  on the subject  \cite{Hall91,HalHaw85,HarHaw83,HaHaHe08,HarWit88,Hawk82,Hawk84}, and of Vilenkin  \cite{Vile86,Vile89}, who proposed a boundary condition for the wave function of the Universe allowing a direct link to inflationary cosmology.

If at the time of Misner's Mixmaster cosmology the main issue was how our Friedmann - Robertson - Walker model could have emerged from such chaotic, anisotropic early stages \cite{BarMat77,CalHu87,DoZeNo67,Matz69,Matz72,MatMis72,Misn68,Paz90,Stew68}, after the inflationary paradigm became dominant that question (if not answered \cite{Calz91,Calz93b,CalSak92,CalSak93}) lost poignancy, since it was accepted that as long as there were one single inflationary patch in the whole Universe that could account for our existence. Thereby the relevant question became not only the properties of the classical Universe emerging from the primordial quantum state of the Universe, but also how such a classical Universe could have emerged at all.

This was the first point of contact between chaos and quantum cosmology, because, as in quantum mechanics more generally, a classical Universe was identified with a Wentzel - Kramers - Brillouin (WKB) solution to the Wheeler-Dewitt equation \cite{Hall91}, and it is known since Einstein that nonintegrability set strong limits to the applicability of the WKB approximation \cite{Eins17}. It soon became clear that if classicality was identified with WKB form, then an Universe could be classical near the cosmological singularity and turn quantum upon expansion, quite the opposite to expectations \cite{CalGon95,CorShe98,LafShe87,Page84}.

In ordinary quantum mechanics, it is generally accepted that properties of a system become actual when they are measured and recorded. This presupposes there are measuring and recording devices external to the system itself, and whose behavior is classical in at least some important respects. Clearly such a framework is inappropriate for cosmology, where the system encompasses the whole Universe. This led Hartle and Gell-Mann to develop from scratch a quantum theory of closed systems, by adopting the consistent histories approach to quantum mechanics \cite{DowHal92,DowKen95,DowKen96,GelHar90a,GelHar90b,GelHar93,GelHar98,Hart92,Hart93a,Hart93b,Hart94,Hart11,Hens09,Kent96,Kent97,Omne94}. This meant to concentrate not on the wave functions of the Universe at a given point in cosmological evolution, but rather on full histories of the Universe and their mutual consistency. 

Same as in quantum mechanics at large, the emergence of a classical state from the quantum maelstrom involves a process of loss of quantum coherence, the so-called decoherence. Since we will take as axiomatic that the quantum mechanics of a closed system is unitary, if decoherence occurs it means the system is not closed. It is interacting with an environment, and by simplicity we shall consider only cases where the environment is actually a physical system on its own right. More abstract settings, for example the environment being high order correlations of the system itself, are conceivable \cite{Anas97,CalHu93,CalHu95a}; we shall not consider them here for pedagogical (and space) reasons. This leaves us squarely within the so-called environment induced decoherence paradigm  \cite{PaHaZu93,PazZur93,PazZur99,Twam93,UnrZur89,Zure81,Zure82,Zure86,Zure91,Zure93,Zure96}.

Actually, environment induced decoherence explains how an ensemble of classical Universes emerges from the quantum state. There is a second part to the question, which is how a single classical Universe is then selected from the ensemble. This process has many points in common with spontaneous symmetry breaking and also with quantum measurement theory, and we regret we have nothing new to offer beyond what is already known in those fields \cite{AlBaNi11}.

The goal of the environment induced decoherence program is not to explain why the density matrix of the system (in our case, the system is the Universe, or rather its projection upon the mini super space) becomes diagonal, since the density matrix is always diagonal in some basis, but rather to point out some preferred basis, and to show that the density matrix inexorably becomes diagonal in that basis. 

The problem of choosing one single preferred basis out of the bewildering array of possibilities (in the consistent histories program there is the equivalent problem of which variables should be monitored as part of the description of the consistent history set) is not particularly new in Science. Max Weber confronted essentially the same problem in trying to find a consistent way of describing capitalism \cite{Webe92}. Weber's answer was that the selected basis should answer to certain internal consistency requirements (in our case, for example, the predictability sieve \cite{PazZur99,Zure02} to be introduced  below provides such a consistency check), but should reflect our interests also - the choice of preferred basis is informed by exactly what we are trying to describe. Weber's criteria are also good for quantum cosmology.

Once we accept that our problem is the emergence of a classical ensemble of Universes with respect to a restricted set of properties chosen a priori, then environment induced decoherence has much to say about the conflict between the semiclassical limit and nonintegrability we have found above. Indeed, not only the validity of the WKB approximation is restored, but also the process of decoherence is largely universal, that is, environment independent \cite{Zure02,Zure03,ZurPaz94,ZuHaPa93}. We shall give a precise description of what we mean below. 

Surprisingly, however, the consequences for quantum cosmology of these insights from environment induced decoherence  have not been worked out yet to our knowledge. This will be the point where we part with the reader, pointing out the new lands we could quickly conquer, if we set to work on it.

This paper is organized as follows. In Section 2 we provide the necessary background on classical and quantum Hamiltonian cosmology, and the consistent histories formulation of the latter. In Section 3 we discuss classical cosmological chaos, both in Bianchi IX and in Friedmann - Robertson - Walker models, and how the fact of classical chaos raises some foundational issues for quantum cosmology. In Section 4 we discuss decoherence in cosmology, by regarding the mini super spaces considered so far as quantum open systems embedded into an environment of gauge invariant primordial fluctuations. We conclude pointing out a few highlights from the literature.

\section{Quantum Cosmology}

\subsection{Hamiltonian Cosmology}

The first step in implementing the quantum cosmology program is to formulate General Relativity in a language adequate for its quantization. This language is the canonical formulation \cite{ArDeMi62,MiThWh73,Ryan72,RyaShe75}.

We begin by writing the so/called ADM decomposition of the interval element

\be
ds^2={g}^{\left(4\right)}_{\mu\nu}dx^{\mu}dx^{\nu}=-{\bar{N}}^2dt^2+g^{\left(3\right)}_{ab}\left(dx^a+\bar{N}^adt\right)\left(dx^b+\bar{N}^bdt\right)
\label{1}
\te
We follow Misner-Thorne-Wheeler (MTW) conventions throughout \cite{MiThWh73}, indexes $\mu$, $\nu$ run from $0$ to $3$, indexes $a$, $b$ run from $1$ to $3$ and $x^0=t$. We also assume natural units with $\hbar=c=1$. $g^{\left(3\right)}_{ab}$ is the induced metric on $t=$ constant surfaces. 

The Einstein - Hilbert action reads

\be
S={m_p^2}\int\:d^4x\:\sqrt{-{g}^{\left(4\right)}\!}{R^{\left(4\right)}\!}
\label{2}
\te
plus  eventually a cosmological constant, $m_p$ is Planck's mass and we have discarded a total divergence. Further integrations by parts allow us to reduce this to

\be
S={m_p^2}\int\:d^4x\:\bar{N}\sqrt{{g}^{\left(3\right)}\!}\left[\bar{K}^{ab}\bar{K}_{ab}-\bar{K}^2+R^{\left(3\right)}\!\right]
\label{2b}
\te
where

\be
\bar{K}_{ab}=\frac1{2\bar{N}}\left[\bar{N}_{a\left|b\right.}+\bar{N}_{b\left|a\right.}-\frac{\partial g^{\left(3\right)}\!_{ab}}{\partial t}\right]
\label{3}
\te
is the extrinsic curvature of the $t=$ constant surfaces
and $\bar{K}=g^{\left(3\right)}\!^{ab}\bar{K}_{ab}$. The symbol $\left|\right.$ denotes a covariant derivative with respect to $g^{\left(3\right)}\!_{ab}$. 

The action \ref{2} does not depend on the time derivatives of the lapse $\bar{N}$ or shift $\bar{N}^a$. These are Lagrange multipliers enforcing the constraints of the theory. The canonical momenta conjugated to $g^{\left(3\right)}\!_{ab}$ are

\be
\pi^{\left(3\right)}\!^{ab}={-m_p^2\sqrt{g^{\left(3\right)}\!}}\left[\bar{K}^{ab}-\bar{K}g^{\left(3\right)}\!^{ab}\right]
\label{4}
\te

The Hamiltonian density is (discarding a total divergence)

\be
H=\bar{N}\bar{H}^0-2\bar{N}_a\bar{H}^a
\label{5}
\te
where

\be
\bar{H}^0=\frac1{m_p^2\sqrt{g^{\left(3\right)}\!}}\left[\pi^{\left(3\right)}\!^{ab}\pi^{\left(3\right)}\!_{ab}-\frac12\pi^{\left(3\right)}\!^2\right]-m_p^2\sqrt{g^{\left(3\right)}\!}R^{\left(3\right)}\!
\label{6}
\te

\be
\bar{H}^a=\pi^{\left(3\right)}\!^{ab}_{\left|b\right.}
\label{7}
\te
$\pi^{\left(3\right)}\!=g^{\left(3\right)}\!_{ab}\pi^{\left(3\right)}\!^{ab}$ and $R^{\left(3\right)}\!$ is the intrinsic scalar curvature computed from $g^{\left(3\right)}\!_{ab}$. The constraints of the theory are $\bar{H}^0=\bar{H}^a=0$; the $\bar{H}^0$ and $\bar{H}^a$ are the generators of coordinate transformations. $H\bar{H}^0$ is the so-called Hamiltonian constraint, while the $\bar{H}^a$ are the momenta constraints.

The space of all three metrics is the so-called super space \cite{DeW70,Fisc70}. The kinetic term in the Hamiltonian $\bar{H}^0$ may be written as proportional to $\left(1/2\right)G^{\left(ab\right)\left(cd\right)}\pi^{\left(3\right)}\!_{ab}\pi^{\left(3\right)}\!_{cd}$ with the super space ``metric''

\be
G^{\left(ab\right)\left(cd\right)}=g^{\left(3\right)}\!^{ac}g^{\left(3\right)}\!^{bd}+g^{\left(3\right)}\!^{ad}g^{\left(3\right)}\!^{bc}-g^{\left(3\right)}\!^{ab}g^{\left(3\right)}\!^{cd}
\label{8}
\te
Point by point, this metric has signature $\left(-+++++\right)$, with the ``time-like'' direction corresponding to the conformal degree of freedom of the three metric, as we shall show presently

\subsection{Singling out the conformal degree of freedom}
A model where matter is represented by a single conformally coupled scalar field is possibly the simplest one which nevertheless is rich enough to allow for the discussion of chaos and decoherence. It will therefore be our model of choice to illustrate the different points to follow. With this in mind, we begin by formulating general relativity in a way where the dynamics of the conformal degree of freedom is singled out.

Let us write the four dimensional metric as ${g}^{\left(4\right)}_{\mu\nu}=a^2\bar{g}_{\mu\nu}$.The three dimensional metric shall be ${g}^{\left(3\right)}\!_{ab}=a^2g_{ab}$. $a=e^{\Omega}$ is a generic scalar field. $\bar{g}_{\mu\nu}$ admits an ADM decomposition as above, with lapse ${\bar{N}}=a{N}$ and shift ${\bar{N}}^a={N}^a$. To make this decomposition unique we should enforce some condition on $g_{ab}$, such as constraining the determinant $g$ constant. After splitting variables in the $3+1$ way and separating out the conformal degree of freedom we get the connection as

\be
{\Gamma}^{\left(4\right)}\!^{\sigma}_{\mu\nu}=\bar{\Gamma}^{\sigma}_{\mu\nu}+\delta^{\sigma}_{\mu}\Omega_{,\nu}+\delta^{\sigma}_{\nu}\Omega_{,\mu}-\bar{g}^{\sigma\lambda}\bar{g}_{\mu\nu}\Omega_{,\lambda}
\te
The Riemann tensor

\bea
{R}^{\left(4\right)}\!^{\rho}_{\sigma\mu\nu}&=&\bar{R}^{\rho}_{\sigma\mu\nu}+
\delta^{\rho}_{\nu}\Omega_{;\sigma\mu}-\bar{g}^{\rho\lambda}\bar{g}_{\sigma\nu}\Omega_{;\lambda\mu}
-\delta^{\rho}_{\mu}\Omega_{;\sigma\nu}+\bar{g}^{\rho\lambda}\bar{g}_{\sigma\mu}\Omega_{;\lambda\nu}\nn
&+&\left(\delta^{\rho}_{\mu}\Omega_{,\nu}-\delta^{\rho}_{\nu}\Omega_{,\mu}\right)\Omega_{,\sigma}
-\left(\delta^{\rho}_{\mu}\bar{g}_{\nu\sigma}-\delta^{\rho}_{\nu}\bar{g}_{\mu\sigma}\right)\bar{g}^{\lambda\tau}\Omega_{,\lambda}\Omega_{,\tau}\nn
&-&\left(\bar{g}_{\sigma\mu}\Omega_{,\nu}-\bar{g}_{\sigma\nu}\Omega_{,\mu}\right)\bar{g}^{\rho\lambda}\Omega_{,\lambda}
\tea
The Ricci tensor

\be
{R}^{\left(4\right)}\!_{\sigma\nu}=\bar{R}_{\sigma\nu}
-\bar{g}_{\sigma\nu}\bar{g}^{\tau\lambda}\Omega_{;\lambda\tau}
-\left(d-2\right)\Omega_{;\sigma\nu}+\left(d-2\right)\left[\Omega_{,\nu}\Omega_{,\sigma}-\bar{g}_{\sigma\nu}\left(\bar{\nabla}\Omega\right)^2\right]
\te
and the scalar curvature

\be
{R}^{\left(4\right)}=\frac1{a^2}\left\{\bar{R}-\left(d-1\right)\left(2\bar{\nabla}^2\Omega+\left(d-2\right)\left(\bar{\nabla}\Omega\right)^2\right)\right\}
\label{concur}
\te
where

\bea
\bar{\nabla}^2\Omega=\frac1{N\sqrt{g}}\nabla_{\mu}\left[g^{\mu\nu}-n^{\mu}n^{\nu}\right]N\sqrt{g}\Omega_{,\nu}\nn
\left(\bar{\nabla}\Omega\right)^2=\left[g^{\mu\nu}-n^{\mu}n^{\nu}\right]\Omega_{,\mu}\Omega_{,\nu}
\tea
Of course, here $d=4$; setting $d=3$ we obtain the formulae appropriate to the intrinsic curvature in the $t=$ constant surfaces.

The 
extrinsic curvature 

\bea
\bar{K}_{ab}&=&a\left\{K_{ab}-\frac1{N}\left[\Omega_{,a}N_{b}+\Omega_{,b}N_{a}+g_{ab}\left(\Omega_{,0}-N^e\Omega_{,e}\right)\right]\right\}\nn
&=&aK_{ab}-\frac1{N}\left[a_{,a}N_{b}+a_{,b}N_{a}+g_{ab}\left(a_{,0}-N^ea_{,e}\right)\right]
\tea
The action

 \bea
S&=&m_p^2\int\:d^4x\:N\sqrt{g}a^2\left\{{R}-\frac13{K}^2+{K}_{ba}{K}^{ba}-\frac 6{N^2}\left(\Omega_{,t}-N^a\Omega_{,a}-\frac{NK}{3}\right)^2\right\}\nn
&+&m_p^2\int\:d^4x\:2N\sqrt{g}\left\{-\mathbf{\Delta}a^2+3g^{ab}a_{,a}a_{,b}
\right\}
\label{confaction}
\tea
The canonical momentum conjugated to $a$ is 

\be
P=-\frac {12m_p^2}Na\sqrt{g}\left[\Omega_{,t}-N^a\Omega_{,a}-\frac{NK}{3}\right]
\te
While the canonical momenta conjugated to the $g_{ab}$ are (recall that $g^{ab}g_{ab,0}=0$)

\be
\pi^{ab}=-a^2m_p^2\sqrt{g}\left[K^{ab}-\frac13g^{ab}K\right]
\te

As expected, the kinetic term associated to the conformal degree of freedom appears with a negative sign.

\subsection{Adding conformal matter}
We now add to the above model a conformal scalar field $\bar{\Phi}$. The action is

\be
S_m=\frac{-1}2\int\:d^4x\:\sqrt{-{g}^{\left(4\right)}}\left\{{g}^{\left(4\right)}\!^{\mu\nu}\bar{\Phi}
_{,\mu}\bar{\Phi}_{,\nu}+\left[m^2+\frac{\left(d-2\right)}{4\left(d-1\right)}{R}^{\left(4\right)}\right]\bar{\Phi}^2\right\}
\te
where $d=4$. We write $\bar{\Phi}=a^{-\left(d-2\right)/2}\Phi$ and proceed to make explicit the conformal degree of freedom as above to get

\bea
S&=&\int\:d^4x\:N\sqrt{g}\left(m_p^2a^2-\frac{\Phi^2}{12}\right)\left[{R}-\frac13{K}^2+{K}_{ba}{K}^{ba}\right]\nn
&-&6m_p^2\int\:d^4x\:N^{-1}\sqrt{g}\left(a_{,t}-N^aa_{,a}-\frac{1}{3}NKa\right)^2\nn
&+&2m_p^2\int\:d^4x\:N\sqrt{g}\left\{-\mathbf{\Delta}a^2+3g^{ab}a_{,a}a_{,b}
\right\}\nn
&+&\frac{1}2\int\:d^4x\:{N}^{-1}\sqrt{g}\left(\Phi_{,t}-N^a\Phi_{,a}-\frac{1}3NK\Phi\right)^2\nn
&-&\frac{1}2\int\:d^4x\:{N}\sqrt{g}\left[g^{ab}\Phi_{,a}\Phi_{,b}+m^2a^2\Phi^2-\frac13\mathbf{\Delta}{\Phi^2}\right]
\label{fullconfaction}
\tea
So now the canonical momenta conjugated to $g_{ab}$ read

\be
\pi^{ab}=-\left(m_p^2a^2-\frac{\Phi^2}{12}\right)\sqrt{g}\left[K^{ab}-\frac13g^{ab}K\right]
\te
The canonical momentum conjugated to $a$ remains the same

\be
P=-\frac {12m_p^2}Na\sqrt{g}\left[\Omega_{,t}-N^a\Omega_{,a}-\frac{NK}{3}\right]
\te
and the field canonical momentum is

\be
p={N}^{-1}\sqrt{g}\left(\Phi_{,t}-N^a\Phi_{,a}-\frac{1}3NK\Phi\right)
\te
leading to the Hamiltonian density

\bea
&&N\pi^{ab}\left[\left(m_p^2a^2-\frac{\Phi^2}{12}\right)\sqrt{g}\right]^{-1}\pi_{ab}+2\pi^{ab}N_{a\left|b\right.}\nn
&-&P\left[\frac N{24m_p^2\sqrt{g}}P-N^aa_{,a}-\frac{N^c_{\left|c\right.}a}{3}\right]\nn
&+&p\left[\frac N{2\sqrt{g}}p+N^a\Phi_{,a}+\frac{1}3N^c_{\left|c\right.}\Phi\right]\nn
&-&N\sqrt{g}\left(m_p^2a^2-\frac{\Phi^2}{12}\right){R}\nn
&-&2m_p^2N\sqrt{g}\left\{-\mathbf{\Delta}a^2+3g^{ab}a_{,a}a_{,b}
\right\}\nn
&+&\frac{1}2{N}\sqrt{g}\left[g^{ab}\Phi_{,a}\Phi_{,b}+m^2a^2\Phi^2-\frac13\mathbf{\Delta}{\Phi^2}\right]
\tea

\subsection{The Wheeler-DeWitt equation}

We proceed now to quantize the theory along the lines of Dirac's quantization for constrained systems \cite{Dira50,Dira58a,Dira64,HaReTe76}.

Leaving aside the subtleties associated with the super space geometry, the basic idea is to associate the super space ``coordinates'' $g_{ab}$ by the operators ``multiplication by $g_{ab}$'', and the momenta $\pi^{ab}$ by the variational derivative with respect to $g_{ab}$. After sorting out operator ordering ambiguities the constraints are mapped into operators. The states of the theory are then the joint null eigenstates of the constraints. Particularly, the Hamiltonian constraint

\be
H_0\Psi=0
\label{9}
\te
takes the form of a time-independent Schr\"odinger equation for the so-called ``wave function of the Universe'' $\Psi$. This is the Wheeler - DeWitt equation \cite{DeW67a,DeW67b,Whee68,Kuch73,Kuch81}.

It is natural to seek an operator ordering such that the Wheeler - DeWitt equation ends up being covariant in super space, if it is understood that $\Psi$ is itself a scalar. If we introduce indexes $A$, $B$ running from $0$ to $5$ to represent all independent pairs of indexes $\left(ab\right)$ and write $x^A\equiv g_{ab}$ then the Wheeler - DeWitt equation becomes a wave equation is super space

\be
H_0\Psi=\left\{\frac1{\sqrt{-G}}\partial_A\sqrt{-G}G^{AB}\partial_B+m_p^4gR+\xi\mathbf{R}\right\}\Psi=0
\label{10}
\te
where $G^{AB}=G^{\left(ab\right)\left(cd\right)}$ is the super space metric, $G<0$ its determinant and $\mathbf{R}$ its curvature. The undetermined constant $\xi$ is a remnant of operator ordering ambiguity. In the following we shall assume ``minimal'' coupling $\xi=0$ (see however \cite{Hall88,Misn72}).

Actually, it is very hard to solve the constraints of the theory in these variables. This has led Ashtekhar and others to take on the problem from a different viewpoint, eventually leading to the loop quantum cosmology program. We shall not discuss these developments here.

\subsection{The Hartle-Hawking and Vilenkin boundary conditions}

Given that the Wheeler - DeWitt equation is analogous to a Schr\"odinger equation is super space, Hartle and Hawking have proposed to solve it by importing the euclidean path integral methods of ordinary quantum theory. Their ansatz reads \cite{Hall91,HalHaw85,HarHaw83,HaHaHe08,HarWit88,Hawk82,Hawk84}

\be
\Psi\left[g_{ab}\right]=\int\:Dg_{\mu\nu}\:e^{-S_E\left[g_{\mu\nu}\right]} 
\label{11}
\te
Here $g_{ab}$ is a three metric defined on some Cauchy surface $\Sigma$. The integral is over all four dimensional non- singular Riemannian manyfolds having $\Sigma$ as their boundary, and over all euclidean metrics on them such that $g_{ab}$ is the induced metric on $\Sigma$. Observe that while the physical space is suppossed to originate from a Big bang, which therefore defines a boundary to the past, the euclidean metrics in the Hartle - Hawking proposal have no boundary other than $\Sigma$. $S_E$ is the Einstein - Hilbert action for the euclidean metric.

The integral in \ref{11} is ill defined because $S_E$ is not nonnegative definite. This is related to the signature of the super space metric $G^{AB}$. One possible way out is to perform a ``second'' Wick rotation (the first one was going from Lorentzian to Riemannian metrics), that is, one integrates over eucliean metrics with complex conformal factors. $\Psi\left[g_{ab}\right]$ then is defined through analytical continuation, and may become complex in the process \cite{Hall91}.

Vilenkin has proposed an alternative to the Hartle - Hawking where the Universe is created \emph{from nothing} through a quantum tunneling process \cite{Vile86,Vile89}. 

\subsection{Mini Super Space models}

Mini super space models are Hamiltonian systems that result from parametrising the metric and  matter fields in the Universe leaving only a finite number of time-dependent but space-homogeneous degrees of freedom indetermined. The parametrisation should be such that introducing the restricted metric in the Einstein - Hilbert action and then taking variations with respect to the parameters should be equivalent to introducing the restricted metric in the Einstein equations themselves. Observe that spatial homogeneity does not mean that the fields are independent of the space coordinates, but only that there is a group of motions which acts transitively on space and leaves the restricted metric invariant \cite{Misn72}.

In a typical mini super space model the parametrization is chosen in such a way that the momenta constraint hold identically at the classical level. If the $q^{\alpha}$ are the independent parameters, then the Einstein - Hilbert action becomes

\be
S=m_p^2\int\:dt\:{N}\:\left[\frac1{2{N}^2}f_{\alpha\beta}\dot{q}^{\alpha}\dot{q}^{\beta}-U\left(q\right)\right]
\label{12}
\te
We have the freedom to choose ${N}$, which could depend on the $q^{\alpha}$. $\dot{q}$ denotes a time derivative, of course, only after we have chosen ${N}$ we will know what kind of time is $t$. 

The canonical momenta are introduced in the usual way

\be
p_{\alpha}=\frac{m_p^2}{\bar{N}}f_{\alpha\beta}\dot{q}^{\beta}
\label{14}
\te
leading to the Hamiltonian constraint

\be
H_0=\frac1{2m_p^2}f^{\alpha\beta}p_{\alpha}p_{\beta}+m_p^2U\left[q\right]=0
\label{15}
\te
The theory is quantized by replacing $q^{\alpha}$ by the operator $q^{\alpha}\cdot$ and $p_{\alpha}$ by $\left(-i\right)\partial/\partial q^{\alpha}$. Operator ordering is chosen in such a way that $-f^{\alpha\beta}p_{\alpha}p_{\beta}$ becomes the D'Alambertian associated to the metric $f_{\alpha\beta}$.

It is clear that simply ignoring the degrees of freedom other than the $q^{\alpha}$ is, at least, a violation of Heisenberg's principle. However, if it can be shown that the $q^{\alpha}$ are in some way the ``slow'' degrees of freedom, then the mini super space may be regarded as the Born - Oppenheimer approximation to quantum cosmology. Eventually, the theory may be improved by including the zero point energy of all other degrees of freedom into $U\left[q\right]$. We shall return to this below.

\subsection{The interpretation of QC}

We shall now discuss different proposals for an interpretation of the Wave Function of the Universe \cite{AndHal93,Hall87,Hall89,Hart83,Hart87,Kief87,Kief07,Vile89}.

Of course, the simplest interpretation is just to assume Born's rule: let $\left|\Psi\right|^2$ be the probability density for three geometries in super space. One obvious problem is that $\Psi$ is not usually square integrable; however there are ways to factor out this divergence and come up with a viable normalization \cite{Hall11}.

A more conceptual problem is the following. We are used and intend to think of cosmology in terms of evolution. We wish statements such as ``the Universe becomes classical when such a such occurs'' to make sense. The Born rule would allow us to say than a smooth isotropic Universe is more or less likely than a foamy fractal one, but gives us no clue about which comes first. Of course there is no ``time'' strictu sensu in quantum cosmology, but the fact itself of the Lorentzian nature of the super space metric suggests there is a causal structure, with well defined Cauchy surfaces arranged in increasing order of some ``Heraclitean'' variable \cite{UnrWal89}. Indeed, for most practical purposes, the volume of a spatial three section, or the average scale factor for open Universes, is a good enough ersatz for ``time''.

The analogy of the Wheeler - DeWitt equation with the Klein - Gordon equation moreover teaches us how to construct a current whose flux over Cauchy surfaces is conserved. This suggests to normalize $\Psi$ to unit total flux, and to consider the normal component of the current as a measure of relative probability within a given Cauchy surface. For the De Sitter mini super space above this is just the ordinary probability current $\left(-i\right)\left(\Psi^*\Psi_{,a}-\Psi\Psi^*_{,a}\right)$. As we can see, it is not nonnegative definite, and therefore the probability interpretation is untenable. Even if we restrict ourselves to ``positive frequency waves'' for which the normal component of the current is nonnegative, the set of allowed wave functions will be different for different Cauchy surfaces, a very unsatisfactory state of affairs.

The drawbacks of these (and other, more sophisticated ones) interpretations of $\Psi$ suggest than something drastic may be called for. Basically, we may take as an observational fact that the Universe became classical as it expanded, and in many examples $\Psi$ adopts a WKB form for large values of the scale factor. If we merge these two insights, we obtain the ``WKB interpretation'' of quantum cosmology, namely, a wave function of the WKB form means that an observer within that Universe will perceive around her a classical Universe evolving under the particular solution of the Einstein equations such that $p=S_{,a}$.  \cite{Hall91}

The situation when there is a coherent superposition of different WKB solutions is much more unclear\cite{Kief88} . It seems some degree of decoherence is necessary to turn such a pure state into an ensemble of WKB alternatives, each of which can then be interpreted of the prediction of a particular classical Universe. The square absolute values of the coefficients in the original superposition give the relative probability of each Universe within the ensemble.

Most importantly, and finally getting squarely into our subject matter, the WKB interpretation will come under fire when the underlying Hamiltonian system is not integrable, because then $\Psi$ may not take a WKB form at all, certainly not for large Universes \cite{CalGon95,CorShe98,LafShe87,Page84}. It is here that we shall see that only a joint consideration of chaos and decoherence yields a viable quantum cosmology

\subsection{Hartle Gell-Mann formalism}

To analyze this question we shall adopt the consistent histories
approach to quantum mechanics\index{consistent histories
approach to quantum mechanics}, in the
version advanced by Gell-Mann and Hartle \cite{DowHal92,DowKen95,DowKen96,GelHar90a,GelHar90b,GelHar93,GelHar98,Hart92,Hart93a,Hart93b,Hart94,Hart11,Hens09,Kent96,Kent97,Omne94}. The idea is to
define a history by a set of projectors $P_{\alpha }$ acting at times $t_{i}$%
. In canonical terms, a history is given by an evolution of the state vector
such that at every time $t_{i}$, it belongs to the proper space of $%
P_{\alpha }\left( t_{i}\right) $. In path integral terms, the
projectors are represented by window functions $w_{\alpha}\left[
x\left( t_{i}\right) \right] $, which take on unit value if the
instantaneous configuration $x$ satisfies the requirements of the
history $\alpha $, and vanish otherwise. The limiting case of a
\textit{fine-grained history}, namely, when $x\left( t\right) $ is
specified for all times, is assigned an amplitude $\exp iS/\hbar $,
as usual in the Feynman path integral formulation. The amplitude for
a \textit{coarse-grained} history defined by window functions
$w_{\alpha}\left[ x\left( t_{i}\right) \right] $ is defined by the
superposition

\begin{equation}
A\left[ \alpha \right] =\int Dx\;e^{iS/\hbar }\psi \left[ x\left(
0\right) \right] \left\{ \prod_i w_\alpha \left[ x\left(
t_i\right) \right] \right\}  \label{l8-1}
\end{equation}
The probability is naturally expressed in terms of a closed time path
integral

\begin{equation}
\mathbf{P}\left[ \alpha \right] =\int DxDx^{\prime }\;e^{i\left[ S-S^{\prime }\right]
/\hbar }\rho \left[ x\left( 0\right) ,x^{\prime }\left( 0\right)
\right] \left\{ \prod_{i} w _{\alpha }\left[ x\left( t_{i}\right)
\right] \right\} \left\{ \prod_{i} w_{\alpha}\left[ x^{\prime
}\left( t_{i}\right) \right] \right\} 
\label{l8-2}
\end{equation}
In this way we may assign a probability to any coarse-grained
history, but these probability assignments are not generally
$consistent$, namely, the probabilities of two mutually exclusive
histories do not generally add up.
Indeed, let us define the \textit{decoherence functional}\index{decoherence functional} of two histories $%
\alpha $ and $\beta $

\begin{equation}
\mathcal{D}\left[ \alpha ,\beta \right] =\int DxDx^{\prime
}\;e^{i\left[ S-S^{\prime }\right] /\hbar }\rho \left[ x\left(
0\right) ,x^{\prime }\left( 0\right) \right] \left\{ \prod_{i}
w_{\alpha}\left[ x\left( t_{i}\right) \right] \right\} \left\{
\prod_{j} w_{\beta }\left[ x^{\prime }\left( t_{j}\right) \right]
\right\}  \label{l8-3}
\end{equation}
$\mathbf{P}\left[ \alpha \right] =\mathcal{D}\left[ \alpha ,\alpha \right] $
but $\mathbf{P}\left[ \alpha \vee \beta \right] =\mathcal{D}\left[ \alpha
,\alpha \right] +\mathcal{D}\left[ \beta ,\beta \right] +2\mathrm{Re}%
\mathcal{D}\left[ \alpha ,\beta \right] \neq \mathbf{P}\left[ \alpha \right]
+\mathbf{P}\left[ \beta \right] $. The probability sum rule $\mathbf{P}%
\left[ \alpha \vee \beta \right] =\mathbf{P}\left[ \alpha \right] +\mathbf{P}%
\left[ \beta \right] $ only applies when the third term vanishes, and in
particular when there is \textit{strong decoherence}, $\mathcal{D}\left[
\alpha ,\beta \right] =0$ for $\alpha \neq \beta $. As physicists, who deal
with reality, we shall be satisfied that a set of mutually exclusive
histories is consistent when $\left| \mathcal{D}\left[ \alpha ,\beta \right]
\right| \ll \mathcal{D}\left[ \alpha ,\alpha \right] ,\mathcal{D}\left[
\beta ,\beta \right] $ whenever $\alpha \neq \beta .$

To a certain extent, the consistent histores approach has the WKB interpretation built in, because the probability of any history will be negligible, at least in the limit $m_p\to\infty$ unless the history contains stationary points of the Einstein-Hilbert action.

A simple set of consistent histories refers to the values of
conserved quantities \cite{HaLaMa95}. First observe that the path
integral expression (\ref{l8-3}) translates into the canonical
expression

\begin{equation}
\mathcal{D}\left[ \alpha ,\beta \right] =Tr\;\left\{ \tilde{T}\left[
\prod_{j}P_{\beta }\left( t_{j}\right) \right] T\left[ \prod_{i}P_{\alpha
}\left( t_{i}\right) \right] \rho \left( 0\right) \right\} .  \label{l8-4}
\end{equation}
The projectors at different times are related in the usual way
$P_{\alpha }\left( t\right) =U\left( t\right) P_{\alpha }\left(
0\right) U^{\dagger }\left( t\right) .$ If a projector commutes with
the Hamiltonian, then it is time-independent, and expression
(\ref{l8-4})  collapses unless all projectors are indeed identical.
The only histories with nonzero probabilities are those defined by
ranges of conserved quantities in the initial state, and they are
automatically consistent if these ranges do not overlap.

\section{Chaos and Cosmology}

The paradigmatic chaotic dynamical system is probably the Baker transformation \cite{Dorf99}. This is a map of the unit square into itself. If $\left(x,y\right)$ is a point in the square, then it is mapped into $\left(2x-\left[2x\right],\left(y+\left[2x\right]\right)/2\right)$, where $\left[\right]$ stands for integer part. Alternatively, if 

\bea
x&=&\sum_{i=0}^{\infty}c_{-i}2^{-i-1}\nn
y&=&\sum_{i=1}^{\infty}c_{i}2^{-i}
\label{31}
\tea
with every $c_i=0$ or $1$. Then the original point is represented by the doubly infinite sequence $c_i=c_i^{\left(0\right)}$, and it is mapped into the sequence $c_i^{\left(1\right)}=c_{i-1}^{\left(0\right)}$. The mapping is obviously time-reversal invariant and area preserving.

There are two typical ways to quantify the chaoticity of the Baker transformation. Suppose we have two points a distance $dx$ apart. Then each iteration will increase this distance by a factor $e^{\lambda}$, where $\lambda=\ln 2$. We say the map has a positive Lyapunov exponent. Since it is area preserving, it must have a negative Lyapunov exponent too, in this case, for points separated along the $y$ direction.

Now suppose we measure the $c^{\left(n\right)}_{0}$ components over a large number of iterations. No matter how many times we do it, we shall always be one bit of information short of predicting the next outcome. We say the map has a Kolmogorov entropy $S_{{K}}=\ln 2$. We notice the Kolmogorov entropy is just the same as the (only) positive Lyapunov exponent. This is a particular instance of the so-called Pesin theorem.

The Baker transformation is discontinuous at the $x=1/2$ line, but this is nonessential. We could as well stretch the square and fold it back into itself without cutting it, obtaining the so-called Smale horseshoe. This way many points in the square are not mapped back into it, but there is an uncountable set of points that belong to the square and remain in it after any number of iterations, either back o forth. Restricted to this set, the dynamics is quite the same as the original Baker.

The amazing thing is that the phase space of nonintegrable Hamiltonian systems generally contains subsets where the dynamics reproduces the Smale horseshoe \cite{ArnAve68,LicLie92,Poin87}. 

\subsection{Chaotic cosmological models}
As we said in the Introduction, mini super spaces are usually restricted to homogeneous space times. That still leaves open the alternative that space may be isotropic (and thus the model belongs to the Friedmann - Robertson - Walker (FRW) class) or anisotropic (thus falling under one of the nine types of the Bianchi classification). We shall be mostly concerned with the first (FRW) case, as it is the most directly relevant to inflationary cosmology. However, nontrivial dynamical behavior was first found in the Bianchi IX type mini super space, and this beautiful problem still is a major motivation in the development of the subject. Therefore, we shall point out some few entry points to the literature on Bianchi IX models before resumming our analysis of FRW ones.

\subsubsection{The Bianchi IX models}
The metric of a homogeneous space time may be written as

\be
ds^2=-\bar{N}^2\left(t\right)dt^2+g_{ab}\left(t\right)\sigma^a\otimes\sigma^b
\label{32}
\te
The $\sigma^a$'s are three independent one-forms obeying

\be
d\sigma^a=C^a_{bc}\sigma^b\wedge\sigma^c
\label{33}
\te
Homogeneity shows in the $C^a_{bc}$ being space-independent. The $C^a_{bc}$ are antisymmetric in $\left(b,c\right)$ but may be of mixed symmetry with respect to $a$. Classifying them according to their symmetry we obtain the nine Bianchi types. The most interesting for our purposes are Bianchi I, where all $C^a_{bc}=0$, and Bianchi IX, where $C^a_{bc}=\epsilon_{abc}$ is totally antisymmetric.

In the Bianchi I case, also known as Kasner Universe, we may simply take $\sigma^a=dx^a$. Then the solution of the vacuum Einstein equations reads

\be
ds^2=-\bar{N}^2\left(t\right)dt^2+t^{2q_1}\left(\sigma^1\right)^2+t^{2q_2}\left(\sigma^2\right)^2+t^{2q_3}\left(\sigma^3\right)^2
\label{34}
\te
The exponents satisfy $q_1+q_2+q_3=q_1^2+q_2^2+q_3^2=1$. They may be parameterised in terms of a single parameter $u$ ranging from $1$ to $\infty$ as

\bea
q_1&=&\frac{-u}{1+u+u^2}\nn
q_2&=&\frac{1+u}{1+u+u^2}\nn
q_3&=&\frac{u+u^2}{1+u+u^2}
\label{35}
\tea
or some permutation thereof.

The Bianchi IX evolution may be understood as a sequence of ``Kasner epochs'', themselves grouped within ``Kasner eras''. In each Kasner epoch, the metric may be approximated by \ref{34} (with the appropriate one forms, of course) and therefore characterized by a parameter $u$ with some well defined value. If $u>2$, then the Kasner epoch eventually evolves into another Kasner epoch with $u'=u-1$. However, if $1<u<2$, then the Kasner era ends, and another era begins with $u'=1/\left(u-1\right)$. 

It follows that if $u_n$ is the value of $u$ at the first epoch of the $n$-th era, then the number of epochs in the era is $\left[u\right]-1$ ($\left[\right]$ standing for integer part). Let us ``compactify'' the $u_n$'s by writing $x_n=u_n^{-1}$. Then $0<x_n<1$, and each $x_n$ is retrieved from the preceding one through the mapping

\be
x_n=\frac1{x_{n-1}}-\left[\frac1{x_{n-1}}\right] 
\label{36}
\te
This mapping has a positive Kolmogorov entropy, and thus it is strongly chaotic. To the best of our knowledge, this insight was the first quantitative indication of chaos in cosmological models \cite{BeKhLi70,Barro82,Misn70}.

The Bianchi IX Universe has been subsequently analysed by a large variety of techniques, confirming the original insight  \cite{BOST01,Barro87,BarSir89,Berg89,Berg90,Berg93b,Burd93,BuBuEl90,BuBuTa91,BurTav92,CoGrRa93,CorLev97a,CorLev97b,Elsk83,FraMat88,FraMat93,HeiUgg09,Hobi91,HBWS91,MBBI08,Misn93,OOST02,Pull91,Rugh93a,Rugh93b,SzyIap90,Zard83}.

We are not aware of a direct proof of Bianchi IX chaos in terms of Ashtekhar variables. Bianchi IX may be regarded as a perturbation of Bianchi VIII, which is explicitly integrable. However, when one tries to carry the computation from Bianchi VIII to Bianchi IX, perturbation theory breaks down. This is of course in agreement with expectations, but falls short of an actual proof of non integrability  \cite{GonTat95,CalThi97}.

\subsubsection{The Friedmann-Robertson-Walker models}

A Hamiltonian system with $n$ degrees of freedom is integrable if it has $n$ constants of motion in involution, which means that all the Poisson brackets between any two of these constants vanish. A FRW mini super space has only one geometrical degree of freedom, the radius of the Universe $a$, and at least one constant of motion, the Hamiltonian itself. Therefore, to be non integrable it must contain non gravitational degrees of freedom, i. e. matter fields, as well. The simplest FRW mini super space model where chaos may be investigated has, besides $a$, a single homogeneous scalar field $\phi$. It is extremely suggestive that already this basic setup leads to nontrivial dynamical behavior  \cite{BoLoCa98,Calz93a,CalElH93}.

The model investigated by Calzetta and El Hasi is spatially closed; the matter field is free, massive and conformally coupled. The model describes a periodic Universe undergoing a sequence of big bangs and collapses. Non integrability becomes apparent when the whole sequence is considered \cite{MotLet02}, rather than a single cosmic cycle. More realistic models including more matter fields already show observable consequences of chaos within a single cosmic cycle  \cite{CalElH95,CorLev96}. See also  \cite{OliSoa99,Toma91}.

By now the consensus is that FRW Universes filled with an scalar field are non integrable except for a few exceptional cases. The field may be real or complex, and conformally  or minimally coupled, and there may be a cosmological constant or not \cite{CoSkSt08,MaWuZh09,MPSS08}

Let us discuss briefly the model analyzed by Calzetta and El Hasi \cite{CalElH93}. It assumes a FRW
spatially closed geometry, that is, a line element

\be
ds^2=a^2(\eta )[-d\eta^2+d\chi^2+\sin^2\chi (d\theta^2+\sin^2
\theta d\varphi^2)]
\te

Where $0\le\varphi\le 2\pi$, $0\le\theta\le\pi$, $0\le\chi\le\pi$,
and $\eta$ stands for ``conformal'' time. For concreteness, we
shall consider only models starting from a cosmic singularity,
that is, we restrict $\eta$ to be positive, with $a(0)=0$.
Also, as indicated by the dynamics, we shall assume that after
the Big Crunch ( that is, when $a$ returns to $0$ ), a new
cosmological cycle begins, now with $a\le 0$. Therefore, a
complete periodic orbit describes the birth and death of two
Universes. The scalar curvature $R=6$ and 
the conformal volume of an spatial surface is $v=2\pi^2$. It is convenient to rescale the conformal factor
$a=a_s/\sqrt{24}\pi m_p$, the homogeneous matter field $\phi=\phi_s/\sqrt{2}\pi$ and the field mass
$m=\sqrt{24}\pi m_p\mu$. Calling $P_s$ and $p_s$ the canonical momenta conjugated to $a_s$ and $\phi_s$, respectively, 
we get the  Hamiltonian

\be
H=({1\over 2})[-(P_s^2+a_s^2)+(p_s^2+\phi_s^2)+\mu^2a_s^2\phi_s^2]\te
The Hamiltonian constraint is $H=0$.  We  introduce the ``adiabatic'' amplitude and phase
$j$ and $\vartheta$

\be\phi_s =\sqrt{{2j\over\omega}}\sin\vartheta\te

\be p_s=\sqrt{2\omega j}\cos\vartheta\te

Where $\omega^2=1+\mu^2a_s^2$ is the instantaneous frequency of the field.
This transformation is canonical if we 
change the geometrical momentum from $P_s$ to $P_1$, according to

\be P_s =P_1+{\mu^2a_sj\over 2(1+\mu^2a_s^2)}\sin 2\vartheta\te
The Hamiltonian becomes $H=-(H_0+\delta H)$, 
where the unperturbed
Hamiltonian

\be
H_0=({1\over 2})[P_1^2+a_s^2]-j\omega\te 
and the perturbation

\be
\delta H=\frac{\mu^2a_sP_1j}{ 2\omega^2}\sin 2\vartheta
+\left[ \frac{\mu^2a_sj}{ 4\omega^2}\right] ^2\left( 1-\cos 4\vartheta \right) \te

The dynamics of $a_s$, as generated by $H_0$, is obviously bounded. The
point $a_s=0$ is a fixed point; it is stable if $\mu^2j\le 1$, , and
unstable otherwise. In this second case, there is an homoclinic loop
associated with it. However, this orbit does not satisfy the
Hamiltonian constraint; rather, we have $H_0=-j$ on the
homoclinic loop.

The equations of motion are simpler if written in terms of
a new variable $X=\omega$, rather than $a_s$ itself. The
transformation is canonical if we associate to $X$ the momentum

\be
P_X=\frac{XP_1}{ \mu\sqrt{X^2-1}}\te
In the large $j$ limit, with $H_0\sim 0$ and $\mu$ fixed.  $X\gg 1$ for most of an orbit, and the Hamiltonian
simplifies to

\be
H_0=(\frac{\mu^2}{ 2})P_X^2+(\frac{1}{ 2\mu^2})[(X-\mu^2j)^2-(\mu^2j)^2-1]
\te
If we parametrize

\be
P_X=\frac{\sqrt{K}}{\mu}\cos 2\alpha\te

\be
X=\mu^2j+\mu\sqrt{K}\sin 2\alpha\te
then

\be
H_0=\left( \frac12\right) [K-(\mu j)^2-\frac{1}{ \mu^2}]\te
We have succeeded
in integrating the unperturbed motion, only that, because the last
reparametrization  involves $j$,
the angle canonically conjugated to it is no longer $\vartheta$,
but

\be\theta =\vartheta -\mu\sqrt{K}\cos 2\alpha\te
The perturbation causes resonances between the variables $\alpha$ and $\theta$ and thus destroys the integrability of the system.
Resonances occur
whenever $\omega_j\pm n\omega_K=0$, 
$\omega_{j,K}$ being the frequencies associated to $\theta$ and $\alpha$,
respectively. We have $\omega_j= \mu^2j$
and $\omega_K
= -1/2$. We thus find a tower of resonances, corresponding to
all positive values of $n$;  the n-th resonance occurs at $j_n= n/2\mu^2$,
independently of $K$. If we further impose the Hamiltonian constraint,
then $K$ must take the value $K_n= \left( \mu j_n\right) ^2=\left( n/2\mu\right) ^2$.

It can be shown that the $n$/th resonance is surrounded by a stochastic layer whose width scales as
$ n^{1/3}/2\mu^2$. For large $n$ each resonance is much wider that the separation
$1/2\mu^2$ between resonances. Thus the stochastic layers merge and the behavior of the perturbed system becomes
chaotic.

\subsection{Chaos and Quantum Cosmology}

Though the ubiquity of chaos in cosmological models is interesting for its own sake, for the purposes of this review the point is how classical chaos impacts on quantum cosmological models. To analyze this issue, we shall show that a WKB solution for small Universes may diverge from the exact solution as the Universe expands, thus making the WKB interpretation of quantum mechanics unviable as is \cite{CalGon95,CorShe98,LafShe87,Page84}.

To see in simple terms why a WKB solution of the Wheeler - DeWitt equation departs from the exact solution when the underlying classical cosmological model is chaotic, let us consider again the procedure whereby the WKB solution is obtained in a multidimensional mini super space model.

Let us return to a generic mini super space model as described in equations \ref{12}, \ref{14} and \ref{15}. Adopting the ''covariant operator ordering, the Wheeler - DeWitt equation becomes

\be
H_0=\left\{\frac{-1}{2m_p^2f^{1/2}}\frac{\partial}{\partial q^{\alpha}}f^{1/2}f^{\alpha\beta}\frac{\partial}{\partial q^{\beta}}+m_p^2U\left[q\right]\right\}\Psi=0
\label{15b}
\te
where $f=\mathrm{det}\:f_{\alpha\beta}$. We regard $\Psi$ as a scalar defined on the mini super space manyfold. We postulate a solution of the form

\be
\Psi=\Delta^{1/2}\:e^{im_p^2S}
\te
We ask that \ref{15b} holds order by order in $m_p^2$. In the limit $m_p^2\to\infty$ the dominant terms are $O\left(m_p^2\right)$ and $O\left(1\right)$. The first yields

\be
\frac{1}{2}f^{\alpha\beta}\frac{\partial S}{\partial q^{\alpha}}\frac{\partial S}{\partial q^{\beta}}+U\left[q\right]=0
\label{15c}
\te
and the second

\be
f^{\alpha\beta}\frac{\partial S}{\partial q^{\alpha}}\frac{\partial}{\partial q^{\beta}}\ln \left[f^{1/2}\Delta\right] +\frac{\partial}{\partial q^{\alpha}}f^{\alpha\beta}\frac{\partial}{\partial q^{\beta}}S=0
\label{15d}
\te
We understand \ref{15c} as meaning the more general equation

\be
\frac{1}{2}f^{\alpha\beta}\frac{\partial S}{\partial q^{\alpha}}\frac{\partial S}{\partial q^{\beta}}+U\left[q\right]=E
\label{15c1}
\te
plus the constraint $E=0$. We recognize \ref{15c1} as the Hamilton-Jacobi equation for the mini super space evolution. The general solution will take the form $S=S\left[q^{\alpha},J_{A}\right]$, where the $J_{A}$ are parameters regarded as world scalars. In general $E=E\left[J_A\right]$, only configurations such that $E=E\left[J_A\right]=0$ are physical.

Observe that \ref{15d} admits the solution \cite{More51,VanV28}

\be
\Delta=f^{-1/2}\mathrm{det}\frac{\partial^2S}{\partial q^{\gamma}\partial J_A}
\label{15i}
\te
To reveal the geometric content of \ref{15i}, observe that we may use $S$ to generate a canonical transformation to new variables

\be
\Theta^A=\frac{\partial S}{\partial J_A}
\te
and then

\be
\Delta=f^{-1/2}\mathrm{det}\left[\frac{\partial q^{\gamma}}{\partial \Theta^A}\right]^{-1}
\label{15j}
\te
The WKB wave function will blow up everytime the matrix inside the determinant becomes singular, that is, at caustics of the classical evolution. In a regular evolution these caustics are isolated; in a chaotic evolution caustics proliferate and the WKB wave function becomes arbitrarily ``spiky''. The exact solutions of the Wheeler - De Witt equation, on the other hand, are regular and cannot reproduce this behavior, therefore the WKB and exact solutions diverge from each other \cite{Berr83,Ozor88}.

This effect has been demonstrated through numerical solutions of the Wheeler - DeWitt equations and their WKB approximations for the conformal field FRW mini super space in refs. \cite{CalGon95} and \cite{CorShe98}.

\section{Decoherence and Cosmology}
\subsection{Environment induced decoherence}
In this review, we shall stick to the position that the quantum mechanics of a closed system is unitary, and that therefore any decoherence effect must be caused through the interaction of the system proper with an environment. As we have remarked in the Introduction, while the formalism itself seems to place no requeriments on which systems qualify as environments, in practice not all choices are equally compelling. 

In the discussion of classical equations from quantum dynamics, Gell-Mann
and Hartle \cite{GelHar93} pointed out that for a large and possibly complex
system the variables that will become classical habitually are the local
densities of conserved hydrodynamic variables integrated over small volumes. To show that some variables become
classical involves showing that they are readily decoherent and that they
obey deterministic evolution equations. Hydrodynamic variables such as
energy, momentum and number are of such characters because they are
conserved quantities \cite{Anas98,Anas01,Anas02,BruHal96,Hall98,Hall03}. In \cite{CalHu99} it is hown that these variables obey deterministic equations in spite of being strongly coupled to the environment because their inertia is correspondingly very high

When such a natural choice is not forthcoming, one possibility is to choose an environment that warrants stability and predictability of the system proper dynamics. This leads to  choose the environment by employing the Paz and Zurek's predictability sieve \cite{PazZur93,Zure02,ZuHaPa93}. The predictability sieve is the most general criterium available for the selection of the preferred basis. It has ben applied to cosmology in \cite{KLPS07} 

For simplicity, we shall only consider cases where system proper and environment are actually two different systems.  We shall only consider histories where the path $\psi^\alpha(t)$ of the system proper is specified, while the path $\xi^\alpha(t)$ of the environment is left unconstrained. If we assume an uncorrelated initial condition, then computing the decoherence functional reduces to a path integral over the environment

\be
 D(\alpha,\alpha')=
\rho_s(\psi_i,\psi'_i)e^{i\left[S_s(\psi^1)-S_s(\psi^2)+S_{IF}\left(\psi^1,\psi^2\right)\right]} 
\ee
where $S_s$ is the free action of the system, $\rho_s$ is the initial density matrix for the system, and $S_{IF}$ is the  Feynman - Vernon influence action \cite{CalLeg83a,CalLeg83b,FeyVer63,FeyHib65,GrScIn88}

\be
e^{iS_{IF}\left(\psi^1,\psi^2\right)}=\int d\xi_i \; d\xi'_i\;d\xi_f\int^{\xi_f}_{\xi_i}D\xi^1
\int^{\xi_f}_{\xi'_i}D\xi^2\;\rho_E(\xi_i,\xi'_i)\;
e^{i[S_E(\xi^1,\psi^1)-S_E(\xi^2,\psi^2)]}
\te
where $S_E$ is the action of environment, including the system-environment interaction and and $\rho_E$ is the environment initial density matrix. 
The weak decoherence condition is recovered when

\begin{equation}
 e^{-Im[S_{IF}(\psi^1,\psi^2)]}<<1\;\;\;\Rightarrow\;\;\;Im[S_{IF}(\psi^1,
\psi^2)]>>1
\end{equation} 
We observe that the influence action $S_{IF}=\Gamma$  is also the Schwinger - Keldysh (or closed time path - CTP) effective action for the environment under the approximation that system variables are treated as c-numbers, that is, Feynman graphs containing system propagators as internal lines are neglected \cite{BakMah63a,BakMah63b,CalHu87,CalHu88,CalHu89,CSHY85,Jord86,Keld64,Schw61,SCYC88,ZSHY80}. This analogy implies that if we seek the effective dynamics for the system by looking at the stationary phase points of the decoherence functional, then this dynamics will be causal and dissipative. On the other hand, we may use the Feynman - Vernon trick of deriving the influence action from coupling the system to a random external source; then we see that the system dynamics is stochastic as well.

The physical basis of dissipation and noise in the system dynamics is discussed in detail in \cite{CalHu08}. What is going on is that the time evolution of the system variables excites the environment through parametric amplification. System and environment become entangled, and thus tracing over the second decoheres the first. The energy for the environment excitation comes from the system, whose own dynamics is therefore damped. The back reaction from the environment on the system has a random component (in the absense of special correlations in the initial state) which is perceived by the system as noise. The unitarity of the underlying theory enforces certain relationships between dissipation and noise, in an equilibrium situation these are just the fluctuation - dissipation theorem. 

We shall now see how the above applies to quantum cosmology.

\subsection{Midi Super Space models}

The basic setup to discuss decoherence in quantum cosmology is to assume the mini superspaces we have studied above are simply the relevant sector of the theory, thus defining everything else in super space as the environment \cite{Hu89,Hu90,Hu91,Hu93,HuPaSi93,HuSin95,PazSin91,PazSin92,SinHu91}. However, since we do not really know how to do quantum cosmology (in the Wheeler - DeWitt sense) in super space, some further simplification is called for. A common approach is to reduce the number of degrees of freedom at the classical level, obtaining a theory with still infinite degrees of freedom but no general dipheomorphism invariance, and thus more amenable to quantization. The resulting theory is called a midi super space. In this review we shall consider midi super spaces consisting of a Friedmann - Robertson - Walker Universe plus homogeneous scalar matter and linearized fluctuations around them.  There are still many midi super spaces answering to this description, as we may consider all scalar, vector and tensorial perturbations, and then further different ways of splitting them into relevant variables and environment. For example, we may consider scalar fluctuations as relevant and tensorial as environmental, or the scalar fluctuation occupation numbers as relevant and the mode phases as environment, or some scalar fluctuations as relevant and the rest as environment, and so on \cite{BrLaMi91,CalGon97,CalHu95a,CaLaLo99,FranCal11,HuKan87,HuKaMa94,HuPav86,HuPaZh93,KiePol09,KiPoSt00,KrOxZa94,LafMat93,LomNac05,Mata93,Mata94,Mori90,MorSas84,Star86}.

A basic discussion of decoherence, dissipation and noise in a FRW minisuperspace coupled to a massive conformally coupled free scalar field is given in \cite{CalHu94}. However in this work linearized gravitational fluctuations were totally neglected, to the effect that the momenta constraint were violated. This may raise doubts about the soundness of the analysis. We shall now show that a more detailed analysis employing so-called gauge invariant variables \cite{AndMuk94,Bard80,FMOV12,GoKoSa10,Mukh05,MuFeBr92,Naka12,Stew90} yields essentially the same results. If anything, the effects of noise and dissipation have been underestimated.

For the sake of discussion we shall consider only the conformal factor of a spatially  closed Friedmann - Robertson - Walker Universe and a homogeneous, conformally coupled massive real scalar field as relevant variables, and the scalar linearized fluctuations around them as the environment. In principle there are four different scalar fields associated to metric fluctuations, as we may perturb the lapse $N=1+A$, the shift $N^a=\gamma^{ab}B_{,b}$, the conformal factor $a\to a+\delta a= a\left(1+\left(1/2\right)\psi\right)$ or the three-metric $g_{ab}=\gamma+E_{\left|ab\right.}-\left(1/3\right)\gamma_{ab}\mathbf{\Delta}E$. There is one more scalar field associated to the matter fluctuations $\Phi=\phi+\varphi$. Here $\gamma_{ab}$ is the natural metric of a three sphere in the corresponding coordinates. We must also decompose the canonical momenta associated to the conformal factor $P=P_0+\delta P$ and matter field $p=p_0+\delta p$. The other geometrodynamic momenta are expressed in terms of a single scalar momentum conjugated to $E$. These make eight degrees of freedom per point.

Neither all of these fluctuations are independent, nor they are physical. Upon a coordinate change $x^{\mu}\to x'^{\mu}=x^{\mu}+\xi^{\mu}$ the metric changes into $g'^{\left(4\right)}_{\mu\nu}=g^{\left(4\right)}_{\mu\nu}-\xi_{\mu;\nu}-\xi_{\nu;\mu}$ and the field into $\Phi'=\Phi-\xi^{\mu}\Phi_{,\mu}$. The corrections to the background quantities are scalar perturbations provided $\xi_a=\xi_{,a}$. A crucial observation is that these ``gauge'' transformations mix the canonical variables with the lapse and shift variables \cite{Teit82}.
Since we have two scalar functions $\xi$ and $\xi_0$ to play with, we may impose two arbitrary conditions. We choose $B=E=0$. If we are given a coordinate system where these conditions do not hold, we may perform a coordinate change as above with $\xi=a^2E/2$ and $\xi_0=a^2\left(B-E_{,t}/2\right)$. This means that instead of the original fluctuations, we take as degrees of freedom the fluctuations in the frame where $B=E=0$. Since this frame is (at least locally) unique, the resulting degrees of freedom are ``gauge invariant''.

The two ``gauge fixing conditions'' $B=E=0$ play formally the role of two new primary constraints on the theory, and so there also bring two secondary constraints meaning $\dot{B}=\dot{E}=0$. Together with the two true secondary constraints $H^0=H^a,_{,a}=0$ we have in total six conditions the linearized fluctuations must satisfy, bringing the number of true degrees of freedom to just two. Moreover, the resulting theory may be brought through a canonical transformation into the theory of a canonical scalar field with a variable mass (see below). The difference with respect to the analysis in \cite{CalHu94} is that the mass depends not only on the conformal factor (as is the case in the classical theory) but also on the homogeneous matter field. In other word, the time-dependent homogeneous matter field and the canonical gauge invariant variable are gravitationally coupled. Therefore, not just the expansion of the Universe, but matter field fluctuations too, may bring forth parametric amplification of the canonical gauge invariant field.

Let us expand on the relationships among the linearized fluctuations. Observe that if $B=E=0$ then $K_{ab}=0$ and then $\pi_{ab}=0$. This in turn implies $E_{,t}=0$, so the gauge condition is consistent. To check whether $\pi_{ab,t}=0$ at the linearized level, we must compute the variation of the Hamiltonian w.r.t. variations $\delta_{ab}$ with $\gamma^{ab}\delta_{ab}=0$. This implies that $g=\gamma$ is not varied. 

Recall that on the background we have $R^a_{bcd}=\delta^a_c\gamma_{bd}-\delta^a_d\gamma_{bc}$, $R_{ab}=2\gamma_{ab}$ and $R=6$, Therefore $\delta R=2\gamma^{bd}\gamma^{ac}\delta_{dc\left|ba\right. }$.
The variation of the Hamiltonian reads

\be
-2\sqrt{\gamma}\left\{\left[N\left(m_p^2a^2-\frac{\Phi^2}{12}\right)\right]_{\left|bc\right.}+N\left(m_p^2a^2-\frac{\Phi^2}{12}\right)_{\left|bc\right.}\right\}_{\mathrm{traceless}}\delta^{bc}
\te
Therefore consistency of the gauge condition in the linearized theory requires

\be
\left(m_p^2a^2-\frac{\phi^2}{12}\right)A+2m_p^2a^2\psi-\frac{\phi}{3}\varphi=0
\te
We must still consider the constraints of the theory. The scalar momentum constraint reads

\be
\left(Pa_{,a}+p\Phi_{,a}\right)-\frac{\gamma}3\left[\gamma^{-1}\left(Pa+p\Phi\right)\right]_{,a}=0
\te
Linearizing we get

\be
P_0a\psi+2p_0\varphi-a\delta P-\phi\delta p=0
\label{linmomconst}
\te
and the Hamiltonian constraint

\bea
&-&\frac 1{24m_p^2\sqrt{\gamma}}P^2+\frac 1{2\sqrt{\gamma}}p^2\nn
&-&6\sqrt{\gamma}\left(m_p^2a^2-\frac{\Phi^2}{12}\right)\nn
&-&2m_p^2\sqrt{\gamma}\left\{-\mathbf{\Delta}a^2+3\gamma^{ab}a_{,a}a_{,b}
\right\}\nn
&+&\frac{1}2\sqrt{\gamma}\left[\gamma^{ab}\Phi_{,a}\Phi_{,b}+m^2a^2\Phi^2-\frac13\mathbf{\Delta}{\Phi^2}\right]=0
\tea
Whose linear part reads

\bea
&-&\frac 1{12m_p^2{\gamma}}P_0\delta P+\frac 1{{\gamma}}p_0\delta p-6m_p^2a^2\psi+{\phi}\varphi+2m_p^2a^2\mathbf{\Delta}\psi\nn
&+&\frac{1}2\left[m^2a^2\left(\phi^2\psi+2\phi\varphi\right)-\frac23\phi\mathbf{\Delta}{\varphi}\right]=0
\label{linhamconst}
\tea
We may use both constraints to express $\psi$ and $\delta P$ as functionals of $\varphi$ and $\delta_p$, or viceversa.
If we make the first choice, then the dynamics of $\varphi$ and $\delta p$ is described by an effective Hamiltonian density

\be
H_{eff}=H-\delta P\delta a_{,t}
\te
We read $\delta a_{,t}$ out of the original Hamilton equations

\be
\delta a_{,t}=-\frac1{12m_p^2\sqrt{\gamma}}\left[\delta P+AP_0\right]
\te
We also expand the scalar linear fluctuations into eigenfunctions of the spatial Laplace operator

\be
\varphi=\sum_{L}f_{L}\left(x\right)\varphi_{L}\left(t\right);\;\delta p=\sqrt{\gamma}\sum_{L}f_{L}\left(x\right)\delta p_{L}\left(t\right)
\te

\be
\mathbf{\Delta}f_{L}=-L^2f_{L}
\te
where ${L}$ represents the integers $L$, $K$, $M$ with $L\ge 0$, $-L\le K,M\le L$, and $L^2=L\left(L+1\right)$. We assume the $f_{L}$ are orthonormal. Then the linearized equations for different modes decouple. Solving the linearized equations for each mode, all perturbations may be expressed in terms of $\varphi_L$ and $\delta p_L$. The effective Hamiltonian  becomes a sum over modes, each mode being a squeezed harmonic oscillator. These may be reduced by standard methods to the Hamiltonian of a harmonic oscillator with a time-dependent instantaneous frequency \cite{CalHu08}. The frequency will depend  on the scale factor, the homogeneous scalar field and their derivatives,  though the latter dependences shall be suppressed by powers of the Planck mass. Therefore not only the expansion of the Universe, but also a time-dependent matter field, will lead to parametric amplification of the fluctuations.

\subsection{Decoherence and particle creation}

We  shall  now  carry  out an analysis of
noise,  fluctuations  and  dissipation  in the minisuperspace as induced by the gauge invariant fluctuations.

As we have seen, each mode of the gauge invariant scalar variable may be reduced to a parametric oscillator.
Quantization 
proceeds by further  decomposing  each  Fourier amplitude in its positive and
negative frequency parts, defined  by  a  suitable  choice of time parameter.
This is accomplished by developing  the  corresponding  mode  on  a  basis of
solutions  of  the  Klein - Gordon  equation,  so
normalized that the positive frequency function has unit Klein - Gordon norm,
the  negative  frequency  function  has  norm  $-1$, and  they  are  mutually
orthogonal in the Klein - Gordon inner product.   Such  a  basis of solutions
constitute a particle model.  Properly normalized particle models are related
to  each other through Bogolubov transformations. Further  identification  of  the
coefficient of the positive  frequency  function  in  the  development of the
field, as a destruction operator,  allows  for the second quantization of the
theory.  The particle model is  also  associated  to a vacuum state, which is
the single common null eigenvector of the  destruction  operators,  and  to a
Fock  basis,  built  from  the  vacuum through the  action  of  the  creation
operators \cite{BirDav82,MukWin07,Park68,Park69,Park71,Park77,Park79,ParTom09}

It is also well known that in a generic  dynamic  space  time,  there  is  no
single  particle  model  which  can be identified outright with the  physical
concept  of  ``particle'';    however, oftentimes it is possible to employ  a
variety  of criteria (such as minimization of the particle number as detected
by a free falling  particle  detector, Hamiltonian diagonalization, conformal
invariance, analytical properties in the euclidean section of the space time,
if any, etc) to single out a preferred particle model in the distant past (or
``IN'' particle model)\cite{CalCas84,Hart83}, and another in the  far  future,  or ``OUT'' particle
model.  In general, these models are not  equivalent, the vacuum of one model
being a multiparticle state in the other. In our problem, the choice of boundary conditions for the path integral above
amounts to a definite choice of the IN particle model,  and  the  IN  quantum
state, in each branch of the closed time path \cite{CalKan93}.   

Since the CTP effective action is independent  of  the OUT quantum states, we
have more freedom in choosing an OUT particle  model.    It  is convenient to
choose a common OUT particle model for both evolutions  (that  is, the Cauchy
data on the matching surface $\eta=\eta^o$ are the same although  the  actual
basis functions  will  be  different).  The positive-frequency time dependent
amplitude functions $f_{\pm}$  for  the  conformal  model  in each branch are
related    to    those    $F$        of        the        OUT     model    by
$f_{\pm}=\alpha_{\pm}F+\beta_{\pm}F^*$        at      $\eta=\eta^o$,    where
$\alpha_{\pm},\beta_{\pm}$ are  the  Bogolubov  coefficients  in each branch,
obeying  the  normalization  condition  $|\alpha_{\pm}|^2-|\beta_{\pm}|^2=1$.
The CTP effective action  is found to be \cite{CalHu94,CalHu08}:

\begin{equation}                \Gamma                =({i\over        2})\ln
[\alpha_-\alpha_+^*-\beta_-\beta_+^*]   \label{Gammabogo}  \end{equation}  
This  expression  is an exact evaluation of the Gaussian path integral defining the influence action.  

If we regard the saddle points of the decoherence functional as the actual classical trajectories, we see that the decoherence functional is essentially the Feynman - Vernon influence action for the minisuperspace variables interacting with the environment provided by the gauge invariant fluctuations; the influence functional  is also closely related to the Schwinger - Keldish effective action. By following through with this analogy, it is easy to show that the resulting equations of motion are real and causal.

The lesson for us is that  there  can be decoherence ($Im \Gamma > 0$) if and
only if there is particle creation in  different  amounts  in each evolution.
(This is also implicit in \cite{PazSin91,PazSin92}.) Indeed, we  can  always  choose the
OUT  model  so  that  $\alpha_+=1$,  $\beta_+=0$,  yielding $\Gamma =(i/2)\ln
\alpha_-$.  The condition for decoherence in this case is  then  $|\alpha_-|>
1$.  But since $|\alpha_-|^2=1+|\beta_-|^2$, this can only happen if there is
particle creation between these two particle models.

To quantify the above considerations, let us return to the FRW plus a massive conformal field mini super space. Let us consider the system evolution to be close to the $n$-th resonance. This means $\omega_j=n/2$, $\omega_K=-1/2$. We assume $n\gg 1$ and $m\le m_p$. Since the field oscillations are much faster than the expansion and contraction of the Universe we may freeze the conformal factor at its peak value. Therefore in the background $a=n/m$ and $P_0=0$, while, with a convenient choice of phase, $\phi=\sqrt{12}\left(m_p/m \right)\sin n\eta$ and  $p=2n\pi^2\sqrt{12}\left(m_p/m \right)\cos n\eta$. Under these conditions the gauge invariant fluctuations become $A=\psi =0$ and $\delta P=\left( 2p_0/a\right)\varphi= 4\pi^2\sqrt{12}m_p\left(\cos n\eta \right)\varphi$. Therefore the $L$-th mode $\varphi_L$ is a periodically driven oscillator with instantaneous  frequency $\Omega^2=\omega_0^2+\omega_1^2\cos 2n\eta$ \cite{CalHu08} with $\omega_0^2=n^2+L^2+L_0^2  $ ($L_0^2=1+8\pi^4$) and $\omega_1^2=8\pi^4$. Because $\omega_0\gg \omega_1$ this is a very narrow resonance and the effects of parametric amplification will not be strong.

However, when we include chaos in our considerations the overall effect is that the instantaneous frequency for each mode acquires a random component. This situation has been studied by Zanchin \emph{et al} \cite{ZMCB98,ZMCB99}, who show that randomness amplifies particle creation in all modes. Therefore we are led to the same conclusion than  \cite{LoCaBo99}, namely, the relevant parameter range for chaos is also the most favourable one for particle creation, and therefore for decoherence.

The physical mechanism
underlying decoherence  in  the  decoherent  history  scheme of Gell-Mann and
Hartle is the same as in the environment-induced scheme
 based on a  reduced  density matrix obtained by 
tracing  over the environmental degrees of  freedom, namely the entanglement of system and environment through particle creation \cite{Calz89,Calz91b,CaCaSc92,CalMaz90,Hu93b,Hu94}. Since the energy of the created particles must be supplied by the minisuperspace variables, it is clear that the back reaction of the environment on the system proper will have a damping effect.  Moreover, quantum fluctuations in the number of created particles result in fluctuations in the back reaction, which are perceived by the system proper as a stochastic driving \cite{CalHu08}. The joint effect of dissipation and noise may induce a qualitative change in the dynamics of the system proper, which is most conspicuous when this system is chaotic \cite{LoCaBo99}. Cosmological models where this stochastic driving is explicitly taken into account are discussed in \cite{Calz99,CaCaVe97,CalHu94,CalHu97,CalVer99,HuVer02,HuVer03,LomMaz97,MarVer99a,MarVer99b,MarVer00}.

\subsection{Chaos, Decoherence and Quantum Cosmology}

In ordinary quantum mechanical systems, the best way to visualize the effect of noise and dissipation on quantum chaos is the Wigner function \cite{Wign32}. Recall the the dynamics of a chaotic system, for example in the neighborhood of an equilibrium state, is exponentially contracting in some directions in phase space, while expanding along others. Up to terms of order $\hbar^2$ the Wigner function obeys the classical Liouville transport equation. Therefore when the system is chaotic, it tends to produce structure in arbitrary short scales along the contracting directions of the dynamics, while smoothing out along the expanding directions. However, when noise and dissipation are factored in, the resulting diffussion introduces a shortest scale where structure may be produced. On the other hand, the expansion of the Wigner function along the unstable directions is limited by the Liapunov exponents, closely related to the Kolmogorov entropy through Pesin's theorem. The combination of these two effects implies that the volume of the set where the Wigner function is substantially non zero increases in time, until it saturates. Since the logarithm of this volume is the thermodynamic entropy, we find that the thermodynamic entropy of the system proper is increasing at a rate which is determined by the Kolmogorov entropy of the system itself. Observe that the rate of entropy increase is largely independent of the system - environment interaction. This simple but compelling picture of the effect of decoherence on quantum chaos has been put forward by Paz and Zurek \cite{ZurPaz94} and since confirmed by detailed numerical simulations \cite{AltHaa12,ALZP95,BarTos11,BiPaSa02,Bona11,GarWis11,GHSS07,HaShZu98,MonPaz00,MonPaz01,ShiHu95}.

To see whether this scenario obtains in quantum cosmology recall that in quantum cosmology we regard the Wave function of the Universe as a scalar field on minisuperspace. The density matrix is therefore akin to a Schwinger function, namely the expectation value of the product of the field evaluated at two different locations. If the minisuperspace is flat, then the Wigner function may be introduced as a Fourier transform of the density matrix with respect to the difference between its arguments. In the general case, however, mini super spaces are described by curved manifolds. In this general case, notwithstanding, a local momentum space may be introduced by using the same techniques as applied to derive quantum kinetic theory from quantum field theory on curved spaces \cite{CaHaHu88,CalHu89b,CalHu08,Habi90,Habi92,HabLaf90}.

The Wigner function will generally obey two equations, a ``mass shell condition'' and a first order dynamic equation. Both equations are covariant in phase space. In the WKB limit, a well defined concept of time arises, namely the time of the classical solution around which the fluctuations are defined. In this limit the dynamic equation becomes equivalent to the usual equations investigated by Paz and others.

Therefore it seems appropriate to conclude that noise and dissipation due to parametric excitation of gauge invariant fluctuations restores the validity of the WKB limit by avoiding the proliferation of spikes in the Wave Function of the Universe, and produce both decoherence and entropy production at a rate determined largely by the mini super space dynamics. However, it is fair to say that this issue has not been explored in the literature with the detail it demands, and therefore our arguments do not intend to be taken as anything but strong hints

\section{Final Remarks}
In this review we have briefly recalled forty something years of efforts towrads the unraveling of the relationships between chaos, decoherence and quantum cosmology. The result notwithstanding seems patchy and inconclused, because some obvious leads have not been thoroughly followed up yet. The most glaring gap, in my view, is whether the translation of the sustantial body of work on  Wigner functions  for classically chaotic systems may hit some snag due to the specifics of reparametrization invariant theories. To the best of my knowledge this issue has not been investigated beyond the semiclassical approximation.

Given the size of the reference section, it may be useful to conclude by pointing out a few highlights; this is of course a very subjective choice on top of the arbitrary selection which created the reference list in the first place.

To begin with, any discussion of quantum cosmology should be set within the framework of our current understanding of the Very Early Universe. Mukhanov's \cite{Mukh05} stands out among many excellent books and is also a very good introduction to the formalism of gauge invariant variables.

Besides the foundational work such as \cite{DeW67a,HarHaw83,Vile86} and \cite{Whee68}, references \cite{Hall91} and \cite{Kief07} are excellent entry points to all aspects of quantum cosmology. Similarly, the books by Giulini \emph{et al}\cite{GJKKSZ03}, Schlosshauer \cite{Schl07} and Weiss \cite{Weis93} are well known introductions to the physics of quantum open systems. The classic exposition of the Gell-Mann/ Hartle formalism is \cite{GelHar90a}.

The foundational papers of cosmological chaos \cite{BeKhLi70} and \cite{Misn70} are undisputed masterpieces. Together with the review \cite{Barro82} they cover much of what is there to be said. For an updated state of the art see \cite{HeiUgg09}.

With regards of chaos in FRW cosmologies, the early paper \cite{CalElH93} already raised many relevant issues. For the state of the art see \cite{CoSkSt08,MaWuZh09,MPSS08}.

The 1993 Kananaskis meeting \cite{HoBuCo93} was arguably a turning point in the development of the subject. Although I have quoted several individual papers from its Proceedings, the volume as a whole has a remarkable sinergia.

To the best of my knowledge, \cite{CalGon95} was the first detailed demonstration of the divergence between solutions of the Wheeler - DeWitt equation in a mini super space and their WKB approximations. See also \cite{CorShe98}.

Our discussion of decoherence in quantum cosmology was essentially the application to the quantum cosmological context of tools which are well proven in the much wider fields of environment induced decoherence and nonequilibrium quantum field theory. For more general introductions to these subjects see \cite{PazZur99} and \cite{CalHu08}, respectively. References \cite{HuPaSi93} and \cite{CalHu94} may be regarded as blueprints for a decoherent quantum cosmology program. For the further progress of the stochastic cosmology program see \cite{HuVer03}.

In spite of the strong theoretical foundations \cite{Berr83} and ground breaking work by Habib and others \cite{Habi90,Habi92,HabLaf90}, the application of Wigner functions to classically chaotic cosmological models is still underdeveloped. I call on my readers to finish up the task.

\section*{Acknowledgments}
This work is supported by the University of Buenos Aires, CONICET and ANPCyT. I thank Maria Isabel Zuleta for  enlightening me  on the connections between sociology and quantum cosmology, and Bei-lok Hu for a collaboration which goes on even when we are not collaborating.

\end{document}